\documentstyle[12pt,aps,prb,floats,eqsecnum,graphicx,psfig,euscript,amsmath]{revtex}

\renewcommand{\max}{\mathop{\rm max}\nolimits}
\def\lapprox{\,\raise0.4ex\hbox{$<$}\kern-0.8em\lower0.7ex\hbox{$\sim$}\,}
\def\gapprox{\,\raise0.4ex\hbox{$>$}\kern-0.8em\lower0.7ex\hbox{$\sim$}\,}

\begin{document}
\bibliographystyle{prsty}
\centerline{\large\bf Goldstone-Mode Relaxation in a Quantized Hall Ferromagnet}
\centerline{\large\bf in the Presence of Smooth Random Potential}
\vskip 3mm
{
\centerline{S. Dickmann}
}

\centerline{\it Institute for Solid State Physics of RAS, Chernogolovka, 142432 Moscow District, Russia}

\vskip 5mm

\begin{abstract}
We discuss the spin relaxation of a strongly correlated
two-dimensional (2D) electron gas (2DEG) in the
quantized Hall regime when the filling factor is close to an odd-integer.
As the initial
state we consider a coherent deviation of the spin system from
the ${\bf B}$ direction and investigate a break-down
of this Goldstone-mode state due to the spin-orbit (SO)
coupling and smooth disorder.
The spin relaxation (SR) process is considered in terms of
annihilation transitions in the system of spin excitons
(magnons).

\end{abstract}
\vskip 4mm

Great bulk of recent SR measurements and theoretical studies
deal with 2D electrons confined in quantum wells or dots.
\cite{do88,zh93,di096,di96,di99,al98,barad92,kh00,me02,kh03}
However, during the last two decades
only a few of works have been devoted to
the SR in the quantized Hall regime proper.\cite{do88,zh93,di096,di96,di99,al98}
Specifically, we imply conditions
under which a 2DEG is in a strong perpendicular magnetic field
($\!B\!\ge\! 10\,$T) and in the absence of holes and magnetic impurities.
We exclude from the consideration the effects of
electrons in the edge states (cf. Ref. \onlinecite{al98}), and restrict
our study to an odd-integer filling.

In the studied problem
the relevant SR
time is actually {\it not a spin dephasing time} but a time of
{\it Zeeman energy relaxation conditioned by a spin-flip process}. Indeed, any spin-flip
means actually a reduction of the Zeeman energy
$|g\mu_{B} B\Delta S_z|$ (${\vec B}\parallel {\hat z}$,
$\Delta S_z=S_z\!-\!S_0$
is the $S_z$ component deviation from the equilibrium value $S_0$, $g\approx -0.44$).
The  mechanism, which makes the relaxation irreversible, has thereby
to provide the energy dissipation. Another necessary condition is
a spin-flip mechanism non-conserving $S_z$.

In the present work the SO coupling is considered as
the cause mixing spin states and the disorder [to be more precise,
the smooth random potential (SRP)] as a dissipation.  A crystal lattice is
still implicitly assumed to be
present as a ``cooler" for 2DEG excitations. We will
suppose that all electron-phonon relaxation processes which do not
change the 2DEG spin state occur much faster than the SR.
[Thermodynamic relaxation
times are estimated to be $\lapprox\!1\,$ns, i.e. well shorter than
the SR time (see the corresponding estimate given at the end of the
paper).]

We thus study the case where the 2DEG is {\it a quantum Hall ``ferromagnet''}
(QHF); i.e. the filling factor is $\nu={\cal N}/N_{\phi}\simeq 2\kappa\!+\!1$,
where ${\cal N}$ and
$N_{\phi}=L^2/2\pi l_B^2$ are the numbers of electrons and
magnetic flux quanta ($L^2$
is the 2DEG area, $l_B$ is the magnetic length). In the high magnetic field
limit, which really represents the solution to the first order in the
ratio $r_c=(e^2/\varepsilon l_B)/\hbar\omega_c$ considered to be small
($\omega_c$ is the cyclotron frequency, $\varepsilon$ is the dielectric
constant), we get the
ground state with zeroth, first, second,... and $(\!\kappa\!\!-\!\!1\!)$-th Landau
levels fully occupied and with
$\kappa$th level filled only by spin-up electrons aligned along ${\bf B}$.
Only a bare handful of experimental
results on the SR in such a QHF were obtained, first indirectly
(the line-widths
of the electron spin resonance (ESR) were measured in Ref. \onlinecite{do88}) and then
directly (in communication of Ref. \onlinecite{zh93} the photoluminescence
dynamics of spin-up and
spin-down states was studied). The measured times are $5-10\,$ns and
exceed by 1-2 orders the SR times observed in quantum wells where
the SR process is governed by the spin interaction with
band holes.\cite{barad92}

As the spin-system perturbation in the QHF we study a coherent
deviation, when the total $S$ number is not changed.
Namely, the initial state is a Goldstone mode
which represents a quantum precession of the vector ${\bf S}$ around
the ${\bf B}$ direction: $|i\rangle\!=\!\left({\hat S}_-\right)^{{}\!N}\!\!|0\rangle$.
Here $|0\rangle$ stands for the QHF ground state and ${\hat S}_-\!=\!
\sum_j\hat{\sigma}^{(j)}_-$ is the lowering spin operator
[$j$ labels
electrons; $\hat{\sigma}_\pm\!=\!(\hat{\sigma}_x\!\pm\! i\hat{\sigma}_y)/2$, where
$\hat{\sigma}_{x,y,z}$ are the Pauli matrices]. The number $N$ is assumed to be
macroscopically large: $0\!\!\ll\!\!N\!\!<\!\!N_{\phi}$. The spin numbers of
the $|i\rangle$
state are $S\!\!=\!\!S_0\!\!=\!\!N_{\phi}/2$ and
$S_z\!\!=\!\!N_{\phi}/2\!\!-\!\!N$
(i.e. $\left.\Delta S_z\right|_{t=0}\!=\!-\!N$).
The total Hamiltonian has the
form
$H_{\mbox{\scriptsize tot}}\!=
\!\sum_j\! H_1^{(j)}\!+\!H_{\mbox{\scriptsize int}}$.
Here $H_{\mbox{\scriptsize int}}$ is the many-electron (Coulomb
interaction) part of the Hamiltonian which has the usual form
(see, e.g., Ref.
\onlinecite{by81}), and $H_1$ is the single-electron part:
$$
  H_1=\hbar^2\hat{{\bf
  q}}^2/2m_e^*-\epsilon_Z\hat{\sigma}_z/2+H_{SO}+\varphi({\bf r})\,,
  \eqno (1)
$$
where
$\hat{{\bf q}}=-i{\bf \nabla}+e{\bf A}/c\hbar$ is a 2D operator,
$\epsilon_Z=|g|\mu_BB$ is the Zeeman energy of one spin-flipped
electron, and $\varphi({\bf r})$ is the SRP field [${\bf r}$
has components $(x,y)$]. The SO
Hamiltonian is specified for the (001) GaAs plane,
$$
  H_{SO}=\alpha\left(\hat{{\bf q}}\times\hat{\mbox{\boldmath $\sigma$}}
  \right)_{\!z}\!+\!
  \beta\left(\vphantom{\left(\hat{{\bf q}}\times\hat{\mbox{\boldmath
  $\sigma$}}
  \right)}\hat{ q}_y\hat{\sigma}_y\!-\!{\hat q}_x\hat{\sigma}_x\right)\,,
                                                            \eqno (2)
$$
and presents a combination of the Rashba term$\,$\cite{by84} (with
the coefficient $\alpha$) and the crystalline anisotropy term$\,$
\cite{dy86} (see also Refs. \onlinecite{di096,di96,di99,kh00}). The
parameters $\alpha$ and $\beta$ are small:
$\alpha,\beta\ll l_B\hbar\omega_c$ (moreover, really
$\alpha\!<\!\beta\!\sim\!10^{-7}\mbox{K}\cdot\mbox{cm}\!<\!l_B\epsilon_Z$).
This enable us to account $H_{SO}$ perturbatively.

If the SRP is assumed to be Gaussian, then it is
determined by the correlator
$K({\bf r})=\langle \varphi({\bf r})\varphi(0)\rangle$,
where $\varphi({\bf r})$ is the SRP field.
We choose also $\langle \varphi({\bf r})\rangle=0$,
i.e. the SRP energy is measured from the center of the Landau
level. In terms of the correlation length $\Lambda$ and
Landau level width $\Delta$, the correlator is
$K({\bf r})=\Delta^2\exp{(-r^2/\Lambda^2)}$.
In the realistic case $\Delta\approx 5-10\,$K, and $\Lambda\sim
30-50\,$nm. We will
study the case
$$
  T\lapprox T^*\ll \epsilon_Z<\Delta\ll e^2/\varepsilon
  l_B<\hbar\omega_c,\quad\mbox{and}
  \quad\Lambda>l_B                                       \eqno (3)
$$
($T$ is the temperature which is actually assumed to be zero in the
calculations; the value $T^*$ will be defined subsequently). All results which follow are obtained in the leading
approximation
corresponding to these inequalities.

The QHF is also remarkable for the following reason: when neglecting
the last two terms in Eq. (1) the spin excitons (SEs) are actually
exact (to the first order in $r_c$) lowest-energy eigen states. The most
adequate description of
the SE states is realized by the SE creation operators.
\cite{di096,di96,di99,dz83} However, before writing them out we choose the
bare single-electron representation. As previously$\,$\cite{di96,di99} we
use the spinor basis which diagonalizes the first three terms in the
Hamiltonian (1)
to the first order in $u=\beta\sqrt{2}/\l_B\hbar\omega_c$ and
$v=\alpha\sqrt{2}/\l_B\hbar\omega_c$:
$$
  \Psi_{\kappa pa}=
  \left( {\psi_{\kappa p}\atop v\sqrt{\kappa\!+\!1}\psi_{\kappa\!+\!1\,p}
  +iu\sqrt{\kappa}\psi_{\kappa\!-\!1\,p}}\right)\,,\quad
  \Psi_{\kappa pb}=\left({-v\sqrt{\kappa}\psi_{\kappa\!-\!1\,p}
  +iu\sqrt{\kappa\!+\!1}\psi_{\kappa\!+\!1\,p}\atop
  \psi_{\kappa p}}\right), \eqno(4)
$$
where $\psi_{\kappa p}=L^{-1/2}e^{ipy}\phi_\kappa (pl_B^2+x)$ is the wave
function of an
electron in the Landau gauge ($\phi_{\kappa}$ is the harmonic
oscillatory
function). We note that here and in the following we present only the
perturbation expansion to within the
framework of the leading order in $u$ and $v$.
The exciton creation operator is
$$
  {\cal Q}_{ab\,{\bf q}}^{\dag}=\frac{1}{\sqrt{ N_{\phi}}}\sum_{p}\,
  e^{-iq_x pl_B^2}
  b_{p+\frac{q_y}{2}}^{\dag}\,a_{p-\frac{q_y}{2}}\,, \eqno (5)
$$
where $a_p$ and $b_p$ are the Fermi annihilation operators corresponding
to the states (4). The annihilation excitonic operator is
${\cal Q}_{ab\,{\bf q}}\!\equiv\! \left({\cal Q}_{ab\,{\bf
q}}^{\dag}\right)^{\dag}\!\equiv\!{\cal Q}_{ba\,-\!{\bf q}}^{\dag}$ and
we employ also the ``shift" operators
${\cal A}_{{\bf q}}^{\dag}\!=
\!{ N_{\phi}}^{-1/2}{\cal Q}_{aa\,{\bf q}}^{\dag}$ and
${\cal B}_{{\bf q}}^{\dag}\!=
\!{ N_{\phi}}^{-1/2}{\cal Q}_{bb\,{\bf q}}^{\dag}$
(${\cal A}_{{\bf q}}\!=\!{\cal A}_{-\!{\bf q}}^{\dag}$,
${}\,{\cal B}_{{\bf q}}\!=\!{\cal B}_{-\!{\bf q}}^{\dag}$.)
In Eq. (5) the orbital index $\kappa$ is dropped since for our
purposes the approximation of projection onto a
single Landau level is quite sufficient. In the following we drop also
the ``spin-orbit" index $ab$ at the operator (5). Such single-level
exciton operators constitute a Lie sub-algebra which is a part of
an Excitonic Representation (ER) algebra (e.g., see Ref. \onlinecite{di02}
and references therein). In our case the relevant commutation rules are
as follows:
$$
  \begin{array}{l}
  \left[{\cal Q}_{\bf q_1},
  {\cal Q}_{\bf q_2}^{+}\right]=
  e^{i\theta_{12}}{\cal A}_{\bf q_1\!-
  \!q_2}{}\!-\!
  e^{-i\theta_{12}}{\cal B}_{\bf q_1\!-
  \!q_2},\quad\mbox{and}\\
  e^{i\theta_{12}}\left[{\cal A}_{\bf q_1},
  {\cal Q}_{{\bf q_2}}\right]\!=
  \!-e^{-i\theta_{12}}
   \left[{\cal B}_{\bf q_1},{\cal Q}_{{\bf q_2}}\right]
  =N_{\phi}^{-1}{\cal Q}_{{\bf q}_1\!+\!{\bf q}_2},
\end{array} \eqno (6)
$$
where $\theta_{12}=l_B^2({\bf q}_1\!\times\!{\bf q}_2)_z/2$. Besides
evidently $\left[{\cal Q}_{{\bf q_1}},
  {\cal Q}_{{\bf q_2}}\right]\!=\!\left[{\cal A}_{{\bf q_1}},
  {\cal B}_{{\bf q_2}}\right]\!=\!0$. The ground state $|0\rangle$ is
completely determined by the equations ${\cal A}_{{\bf q}}|0\rangle\!=
\!\delta_{{\bf q},0}|0\rangle$ and ${\cal B}_{{\bf q}}|0\rangle\!=\!0$.
Single-exciton states are normalized: $\langle 0|{\cal Q}_{{\bf q}_1}
{\cal Q}_{{\bf q}_2}^{\dag}|0\rangle\!=\!\delta_{{\bf q}_1,{\bf q}_2}$.

In the limit $\Delta \to 0,\;\; H_{SO}\to 0$ and at $r_c\ll 1$ the state
$$
  \left|N;1;{\bf q}\right\rangle\!=
  \!{\cal Q}_{\bf q}^{\dag}\!
  \left({\cal Q}_{0}^{\dag}\right)^N\!\!\left|0\right\rangle  \eqno (7)
$$
is the eigen state of the system studied. It has the spin numbers
$S=N_{\phi}/2\!-\!1$ and $S_z=N_{\phi}/2\!-\!1\!-\!N$ [see below the
expressions (8) which
should be used to calculate $S$ and $S_z$]. The corresponding energy is
$(N\!+\!1)\epsilon_Z+{\cal E}_q$, where ${\cal E}_q$ is the exchange part of
the SE energy. The small momentum approximation
${q}l_B\!\ll\! 1$ is quite sufficient for our problem, and therefore
${\cal E}_q\!=\!(q\l_B)^2/2M_{\mbox{\scriptsize  x},\kappa}$
(general expressions for the
2D magneto-excitons can be found in Ref.
\onlinecite{by81}). Here
$M_{\mbox{\scriptsize  x},\kappa}$ is the SE mass at $\nu=2\kappa\!+\!1$, namely:
$1/M_{\mbox{\scriptsize  x},0}\!=\!(e^2/\varepsilon
l_B)\sqrt{\pi/8}$, $1/M_{\mbox{\scriptsize  x},1}\!=\!7/4M_{\mbox{\scriptsize  x},0}$,...

Spin operators in terms of the ER are
invariant with respect to $H_{SO}$:
$$
  {\hat S}_z=N_{\phi}\left({\cal A}_0-{\cal B}_0\right)/2,\quad
  {\hat S}_-={N_{\phi}\!}^{1/2}
  {\cal Q}_0^{\dag},\quad {\hat {\bf S}}^2=N_{\phi}{\cal Q}_0^{\dag}{\cal
  Q}_0+{\hat S}_z^2+{\hat S}_z\,, \eqno (8)
$$
At the same time in the basis (4) the operators $\varphi({\bf r})$ and
$H_{\mbox{\scriptsize int}}$ acquire
corrections proportional to $u$ and $v$.
Specifically, calculating $\int \Psi^{\dag}\varphi({\bf r})\Psi d^2{\bf r}$,
where
$\Psi\!=\!\sum_p(a_p\Psi_{\kappa pa}\!+\!b_p\Psi_{\kappa pb})$, we get
the terms responsible for a spin-flip process:
$$
  \hat{\varphi}=N_{\phi}^{1/2}l_B\sum_{\bf q}
  \overline{\varphi}
  ({\bf q})\left(iuq_+-vq_-\right)
  {\cal Q}_{\bf q}+\mbox{H.c.}\quad (\mbox{at}\;\;\, ql_B\ll 1)\,.  \eqno (9)
$$
Here $\overline{\varphi}({\bf q})$ is the Fourier component
[i.e. $\varphi\!=\!\sum_{\bf q}\overline{\varphi}({\bf q})e^{i{\bf qr}}$],
and $q_\pm\!=\!\mp i(q_x\pm iq_y)/\sqrt{2}$.

We stress one essential feature of the states (7). In spite of the existence
of a formal operator
equivalence ${\cal Q}^{\dag}_0\equiv
\lim\limits_{q\to 0}{\cal Q}^{\dag}_{\bf q}$
we find that $|N;1;0\rangle$ and
$\lim\limits_{q\to 0}|N;1;{\bf q}\rangle$ present {\it different} states.
Indeed,
in these states the system has different total spin numbers $S=N_{\phi}/2$
and $S=N_{\phi}/2\!-\!1$ respectively.
[$S_z\!=\!N_{\phi}/2\!-\!1\!-\!N$ is the same for both.] So, the excitation
of a ``zero" exciton (with zero 2D
momentum) corresponds to the transition $S_z\!\to\!S_z\!-\!1$ without any
change of
the total $S$ number, but each ``nonzero" SE changes the spin numbers by 1:
$S\!\to\!S\!-\!1$, $S_z\!\to\!S_z\!-\!1$. Let us introduce the notation
$|N\rangle\!=\!\left({\cal Q}_{0}^{\dag}\right)^N\!\!\left|0\right\rangle$.
The initial state $|i\rangle$  in the ER is actually a
``Goldstone condensate" (GC) containing
$N$ zero excitons: ${}\!|i\rangle\!=\!|N\rangle$.

Our goal is to study the process of the GC break-down.
The state $|N\rangle$
is certainly degenerate and we will solve the problem in terms of the
quantum-system transitions within a continuous spectrum. The transition
probability is determined by the Fermi Golden Rule:
$w_{fi}=(2\pi/\hbar)|{\cal M}_{fi}|^2\delta(E_f-E_i)$. In our case
evidently the final state $|f\rangle$ is obviously the state where a part
of the Zeeman energy has
been converted into an exchange energy. Since it is exactly
the single-electron terms that constitute the perturbation responsible for
the ${\cal M}_{fi}$ matrix element, we could find that such a transition is
the $2X_0\!\!\to\!\!X_{{\bf q}^*}$ process in the lowmost
order of the perturbative approach (we denote the zero exciton by $X_0$ and
the nonzero one by $X_{\bf q}$). In other words the final state for this
transition is $|f\rangle\!=\!|N\!-\!2;1;{\bf q}^*\rangle$.
The value $q^*$ is determined by the energy conservation equation $E_f\!=\!E_i$
which reads $2\epsilon_Z\!=\!\epsilon_Z\!+\!{\cal E}(q^*)$, i.e.
$q^*=\sqrt{2M_{\mbox{\scriptsize  x},\kappa}\epsilon_Z}/l_B$.

We point out that the SO interaction (2) alone does not provide a quantum
fluctuation from the GC to any state
with a different electron density (in particular, to the state $|N\!-\!2;1;{\bf
q}^*\rangle$). This feature may be verified by means of direct analysis
of the SO influence on the 2DEG spectrum (c.f. Ref.
\onlinecite{di96}). Besides, this can be recognized from general
considerations. Indeed, each of the
components ${\hat q}_i$ ($i\!=\!x,y$) commutes with any component of
the operator
${\hat {\bf P}}\!=
\!\sum_j [\hbar{\hat{\bf q}}_j-(e/c){\bf B}\!\times\!{\bf r}_j]$.
The latter commutes with the Hamiltonian $\sum_j\hbar^2{\hat{\bf
q}}^2_j/2m_e^*\!+\!
H_{\mbox{\scriptsize int}}$ and one can also check that
$\left[{\hat{\bf P}}, {\cal
Q}^{\dag}_{\bf q}\right]\!=\!q{\cal Q}^{\dag}_{\bf q}$
[if ${\cal Q}^{\dag}_{\bf q}$ is defined by Eq. (5) to within the
zero order in $H_{SO}$].
The operator ${\hat {\bf P}}$ hence plays
the role of 2D momentum in the magnetic field.\cite{go68} Its
permutability with the total 2DEG Hamiltonian
reflects the spatial homogeneity of the system under consideration in
the ``clean limit".
Since $[H_{SO},{\hat{\bf P}}]\!\equiv\! 0$, the SO coupling by itself cannot
destroy the homogeneity in any order of the perturbation approach.

The transition is thereby determined by the operator (9). The
corresponding matrix element
${\cal M}_{if}\!=\!\langle\!\langle {\bf q}^*;1;N\!-\!2|\!|
\hat{\varphi}|\!|N\rangle\!\rangle[R(N)R(N\!-\!2;1;{\bf q}^*)]^{-1/2}$
was actually calculated in Ref.
\onlinecite{di96}. (The SRP plays the same role as the phonon field studied
there.) Here and in the following the notation $R(...)$ stands for the
norm of the state $|...\rangle$.\cite{dz83} The result is
$
  \left|{\cal M}_{if}\right|^2=\frac{N(N\!-\!1)}{2N_{\phi}}(u^2\!+\!v^2)
  \left|q^*l_B\overline{\varphi}({\bf q}^*)\right|^2\,,
$
and we obtain the rate of the
$i\!\to\!f$ transition
$
  (2\pi/\hbar)
  \sum_{\bf q}
  \left|{\cal M}_{if}\right|^2\delta(q^2l_B^2/2M_{\mbox{\scriptsize  x},\kappa}\!-\!\epsilon_Z)\!=\!
  N(N\!-\!1)/\tau N_{\phi}\quad (N\ge 1) ,
$
where
$$
  1/\tau={8\pi^2(\alpha^2\!+\!\beta^2)M_{\mbox{\scriptsize  x},\kappa}^2\epsilon_Z
  {\overline K}(q^*)}/{\hbar^3\omega_c^2 l_B^4}\,.  \eqno (10)
$$
Here ${\overline K}$ stands for the Fourier component of the
correlator: ${\overline K}(q)\!=\!L^2\left|{\overline
\varphi}(q)\right|^2\!/4\pi^2$.

The quantum transition to the state $|f\rangle\!=\!|N\!-\!2;1,{\bf q}^*\rangle$
is certainly a first step in the SR process. This state is thermodynamically
unstable. In a time which is much shorter than $\tau$ it turns to a state
$\!|N\!-\!2;1,{\bf q}_0\rangle$, where $q_0$ takes the lowest possible
nonzero value. In fact, relevant values of $q_0$ are determined by the SRP
field. The SE interaction with the SRP incorporates
the energy $U_{\mbox{{\scriptsize x-}{\tiny SRP}}}\!\sim\!
ql_B^2\Delta/\Lambda$ (the ``nonzero" SE possesses the dipole momentum
$el_B^2[{\bf q}\!\times\!{\hat z}]$, see Ref. \onlinecite{by81}). The latter
determines the inhomogeneous uncertainty of the SE momentum
$\delta q\!\sim\! M_{\mbox{\scriptsize  x},\kappa}\Delta/\Lambda$
(this follows from
the equation $\delta q\partial{\cal E}_{\bf q}/\partial q\!=\!
U_{\mbox{{\scriptsize x-}{\tiny SRP}}}$). Therefore ``quasi-zero" wave
numbers in the range defined by inequalities
$$
  0<q_0\lapprox M_{\mbox{\scriptsize  x},\kappa}\Delta/\Lambda\,,
  \eqno (11)
$$
present the lowest physical limit for momenta of the nonzero SEs.
Under the conditions (3) we find that $q_0\!\ll\!q^*$, and the SE exchange
energy
$(q_0l_B)^2\!/2M_{\mbox{\scriptsize  x},\kappa}$ is smaller than the
value $T^*\!\!=\!\!M_{\mbox{\scriptsize  x},\kappa}\!(\Delta l_B/\Lambda)^2$.
This in turn is negligible in comparison with $\epsilon_Z$.

To solve the problem in a complete form we have
obviously to study the general state of the type:
$$
  |N;M_1,M_2,...,M_K\rangle=\left({\cal Q}_{{\bf q}_{01}}^{\dag}\right)^{M_1}
  \!\!\left({\cal Q}_{{\bf q}_{02}}^{\dag}\right)^{M_2}\!\!...
  \left({\cal Q}_{{\bf q}_{0K}}^{\dag}\right)^{M_K}\!|N\rangle.  \eqno (12)
$$
All the wave-vectors $q_{0k}$ are assumed to satisfy the condition (11).
We will also use for this
state a shorthand notation $|N;M\rangle$,
where $M\!=\!\sum_k^K M_k$ is the total number of the nonzero SEs.
If $M\!\gg\! 1$, we assume that $1\ll K\ll N_{\phi}$.
In the framework of our approach the state (12) is an approximate
eigen state of a QHF having energy $(N\!+\!M)\epsilon_Z$ and spin numbers
$S_z\!=\!N_{\phi}/2\!-\!N\!-\!M$ and $S\!=\!N_{\phi}/2\!-\!M$. This
value of $S_z$ is the exact one. It can be calculated employing the
representation of Eqs. (8) and the commutation rules (6). The same algebra
allows us to find that ${\hat{\bf S}}|N;M\rangle\!=
\!S(S\!+\!1)\left(|N;M\rangle\!+\!|
{\tilde{\varepsilon}}\rangle\right)$, where the norm of the state
$|{\tilde{\varepsilon}}\rangle$ is small compared with the norm of
$|N;M\rangle$, namely:
$R({\tilde{\varepsilon}})/
R(N;M)\!=\!O(m^3n/K)$. The notations
$n\!=\!N/N_{\phi}$ and $m\!=\!M/N_{\phi}$
are used for ``reduced" quantum numbers.
In the special
case when $N=0$, the state $|0,M\rangle$ can be treated as a
``thermodynamic condensate" (TDC) which arises if $M$ is larger
than the
critical number of nonzero SEs. The latter is estimated at
$N_{\phi}l_B^2\!\int\!\frac{d^2\!{\bf q}}{2\pi}/
\!\{\exp{[({\cal E}_q\!+
\!|U_{\mbox{{\scriptsize x-}{\tiny SRP}}}|)/T]}-\!1\}$
(e.g., c.f. Ref. \onlinecite{di96}), and in our case (3) it is at least
smaller than $N_{\phi}M_{\mbox{\scriptsize  x},\kappa}T$. At the same time, $M$ is determined by
the spin $S$ of the system,
therefore at a {\it given} $M\!\!=\!\!N_{\phi}/2\!\!-\!\!S\!\!>\!\!
 N_{\phi}M_{\mbox{\scriptsize  x},\kappa}T$ we find that below some
threshold temperature
the nonzero SEs necessarily form a TDC.
For macroscopically large $N$ and $M$
the state (12) hence features a coexistence of GC and TDC. It should also be
noted that specific values ${\bf q}_{0k}$ as well as specific
distribution given by $M_{k}$ numbers
have no physical meaning. The final
results should not depend on them but only on $M$ and $N$.

We can now write the kinetic equations corresponding to the relevant spin
transitions. The rate $dN/dt$ is determined by the $2X_0\!\!\to\!\!
X_{{\bf q}^*}(\to\!\!X_{{\bf q}_0})$ process (which presents a GC
depletion with a simultaneous ``flow" to TDC)
and by the $X_0\!+\!X_{{\bf q}_0}\!\!\to\!\!
X_{{\bf q}^*}(\to\!\!X_{{\bf q}'_0})$ one. The rate $dM/dt$ is also formed
by the $2X_0\!\!\to\!\!X_{{\bf q}^*}(\to\!\!X_{{\bf q}_0})$
transition (which provides a TDC evolution) and by the $X_{{\bf q}_0}\!+\!X_{{\bf q}'_0}\!\!\to\!\!
X_{{\bf q}^*}(\to\!\!X_{{\bf q}''_0})$ one (determining a TDC depletion).
[Values of $q_0,q_0'$ and $q_0''$
belong to the region (11).] The corresponding equations are derived again
with help of the Fermi Golden Rule and Eqs. (6)
(with vanishing $\theta_{12}$ in the latter):
$$
  dn/dt=-(2\mu_{nn}+\mu_{nm})/\tau,\quad\mbox{and}\quad
  dm/dt=(\mu_{nn}-\mu_{mm})/\tau\,, \eqno (13)
$$
where
$$
  \begin{array}{lcl}
  \displaystyle{
  \mu_{nn}}&\displaystyle{=}&\displaystyle{\frac{|\langle M;N\!-\!2|{\cal Q}_{{\bf q}^*} {\cal Q}_{-\!{\bf q}^*}
  |N;M\rangle|^2}{R(N;M)R(N\!-\!2;M\!+\!1;{\bf q}^*)}}=
  \displaystyle{\frac{N^4R(N\!-\!2;M\!+\!1;{\bf
  q}^*)}{N_{\phi}^2R(N;M)}
  \left[1\!+\!O\left(\frac{m}{nN_{\phi}}\right)\right]},\\
  \displaystyle{
  \mu_{nm}}&\displaystyle{{}{}=}&{}
  \displaystyle{\sum_{k}\frac{|\langle
  M_1,...M_k\!-\!1,...M_K;N\!-\!1|{\cal Q}_{{\bf q}^*\!+\!{\bf q}_{0k}}
  {\cal Q}_{-\!{\bf q}^*}|
  N;M\rangle|^2}{R(N;M)R(N\!-\!1;M;{\bf q}^*)}}\\
  {}\,{}&{}\qquad{}&\displaystyle{=}\quad\displaystyle{
  \frac{4M^2N^2R(N\!-\!1;M;{\bf q}^*)}{N_{\phi}^2R(N;M)}
  \left[1\!+\!O\left({K}/{N_{\phi}}\right)\right]},\\
\displaystyle{
  \mu_{mm}}&{}\displaystyle{=}&\displaystyle{\sum_{k<i}
  \frac{|\langle
  M_1,...M_k\!-\!1,...M_i\!-\!1,...M_K;N|{\cal Q}_{{\bf q}^*\!+\!{\bf q}_{0k}
  \!+\!{\bf q}_{0i}}
  {\cal Q}_{-\!{\bf q}^*}|
  N;M\rangle|^2}{R(N;M)R(N;M\!-\!1;{\bf q}^*)}}\\
  {}\,{}&{}\qquad{}&\displaystyle{=}\quad\displaystyle{
  \frac{2M^4R(N;M\!-\!1;{\bf q}^*)}{N_{\phi}^2R(N;M)}
  \left[1\!+\!O\left({K}/{N_{\phi}}\right)\right]}
  \end{array}
$$
[$R(N;M\!+\!1;{\bf q}^*)$ is the norm of the
${\cal Q}^{\dag}_{{\bf q}^*}|N;M\rangle$ state]. In this way we find
that the norms appearing in these equations satisfy the conditions
$R(N;M\!+\!1;{\bf q}^*)/R(N;M)=r$, $R(N\!+\!1;M)/R(N;M)=Nr_n$ and
$R(N;M_1,...M_k\!\!+\!\!1,...M_K)/R(N;M)=M_kr_m$,
so that
$$
  \mu_{nn}={n^2r}/{r_n^2},\quad \mu_{nm}={4mnr}/{r_nr_m},\quad
  \mu_{mm}=2{m^2r}/{r_m^2},   \eqno (14)
$$
where the factors $r,r_n,r_m$ are determined by the equations
$$
  \begin{array}{l}
  {1=(1-n-2m)/{r_n}+O(m^3/nK),\quad 1=(1-2n-2m)/{r_m}+
  {n^2}/{r_n^2}+O(m^2)},\\
  {1=(1-2n-2m)/{r}+{4mn}/{r_nr_m}+{n^2}/{r_n^2}+
  {2m^2}/{r_m^2}+O(1/N_{\phi})}\,.
  \end{array}
  \eqno (15)
$$
(Positivity of $r_n\!\approx\! 1\!-\!n\!-\!2m$
provides the physically obvious requirement $|S_z|\!<\!S$.)
The last terms in Eqs. (15) would just depend on
the specific set of the $M_k$ numbers. We can therefore calculate $r_m$ and
obtain the final result only in the $m\!\ll\!1$ case.
Meanwhile, the values of $m(t)$ are determined by the
initial value $n(0)$. According to Eqs. (13)-(15)
$\max(m)\!\approx\!n(t^*)[1\!-\!n(t^*)]/\sqrt{2}$ ($t^*$ is the time at
which $m$ peaks), i.e. at least $m^2\!<\!1/32$.
[In particular, at $n(0)\!\!=\!\!0.5$ $\max(m)\!\!\approx\!\!0.1$.]
The problem (13)-(15) is thus solved to the leading order in $m$. We
should put
$r\!\!=\!\!r_n\!\!=\!\!r_m\!\!=\!\!1$ in $\mu_{mn}$ and $\mu_{mm}$, but
$r_n\!\!=\!\!1\!-\!n$
and $r\!\!=\!\!(1\!\!-\!\!n)^2$ in $\mu_{nn}$.
This yields the analytical result
$n(t)\!=\!1/{[2n(0)(t/\tau)^2\!+\!2(t/\tau)\!+\!1/n(0)]}$ and
$m=n(t)n(0)t/\tau$.

The dependences are shown in Fig. 1. The vector
${\bf S}(t)$  at moments $t=0,\tau,2\tau,...$
is depicted in the inset. The time (10) hence
governs the breakdown of the
GC. (Under the realistic conditions above,
$\tau\!\sim\!10^{-8}-10^{-7}\,$s.) However the SR occurs
certainly non-exponentially
and the actual time is increased by a factor of $N_{\phi}/\Delta
S_z(0)$.

Experimentally, the GC could probably be created by microwave
pumping at the electron frequency $\epsilon_Z/\hbar$. This should cause to
``rotate"
the QHF spin without changing of the ${\bf S}$ modulus. As to
observing of
the SR variation with time, one can think that the optical
technique{}\,\cite{zh93} is relevant in the case.

I acknowledge
support by the MINERVA Foundation and by the Russian Foundation
for Basic Research. I thank also the Max Planck Institute for Physics
of Complex Systems (Dresden) and the Weizmann Institute of
Science (Rehovot) for hospitality.

\begin{figure}

\centerline{\psfig{figure=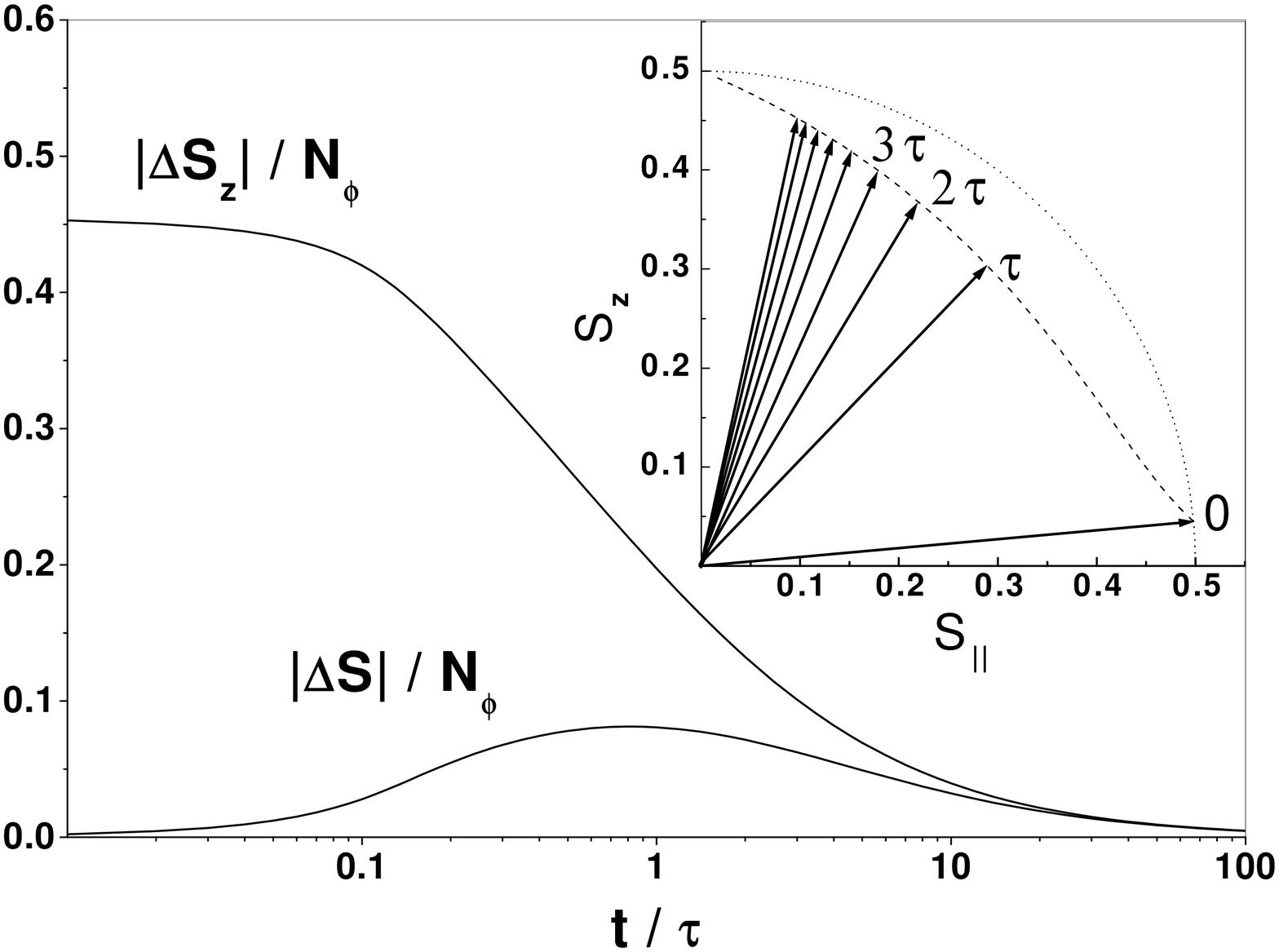,width=150mm,angle=0}}
\caption{Time dependences of
$|\Delta S_z|/N_{\phi}\!\!=\!\!n(t)\!\!+\!\!m(t)$
and of $|\Delta S|/N_{\phi}\!\!=\!\!m(t)$
are shown in the main picture for
$n(0)\!\!=\!\!|\Delta S_z(0)|/N_{\phi}
\!\!=\!\!0.455$. The vectors ${\bf S}(t)$
at equidistant moments of time are plotted in the inset
with step $\tau$. The dotted line is the
arc with radius
$S_0\!\!=\!\!N_{\phi}/2$. The gap between the dashed and dotted lines
reflects the deviation of the spin modulus from the value of $S_0$.}

\label{Firstfig}
\end{figure}

\end{document}